\documentclass[twocolumn]{aastex62}

\usepackage{newtxtext,newtxmath}

\usepackage[T1]{fontenc}
\usepackage{ae,aecompl}

\usepackage{amsfonts,amsmath,amstext,amsgen,amsopn,amsxtra,indentfirst,times}%

\usepackage{savesym}%
\usepackage{graphicx}%
\usepackage{hyperref}%
\usepackage{lineno}%
\usepackage{mathtools}%
\usepackage{multirow}
\usepackage{amsmath}%
\usepackage{xspace}%
\usepackage{units}%
\usepackage{tikz}%
\usetikzlibrary{positioning}%
\usetikzlibrary{shapes,arrows}%
\usepackage{nicefrac}%
\usepackage{array}

\usepackage[version=4]{mhchem}

\newcommand{\RE}{R$_{\rm \earth}$\xspace}
\newcommand{\ME}{M$_{\rm \earth}$\xspace}

\newcommand{\piso}{$P_{\rm iso}$\xspace}
\newcommand{\wmf}{$x_{\rm H2O}$\xspace}
\newcommand{\lupi}{$v^2$ Lupi b\xspace}

\newcommand{\rev[1]}{\textcolor{black}{#1}} 
\newcommand{\cdnote[1]}{\textcolor{black}{#1}} 
\newcommand{\tlnote[1]}{\textcolor{black}{#1}} 
\newcommand{\rgl}{runaway greenhouse limit\xspace}
\newcommand{\rgt}{runaway greenhouse threshold\xspace}

\shorttitle{Hidden water in magma oceans}
\shortauthors{Dorn \& Lichtenberg}

\begin{document}

\title{Hidden water in magma ocean exoplanets}

\correspondingauthor{Caroline Dorn}
\email{cdorn@physik.uzh.ch}

\author[0000-0001-6110-4610]{Caroline Dorn}
\affil{University of Zurich, Institute for Computational Science, Winterthurerstrasse 190, CH-8057, Zurich, Switzerland}

\author[0000-0002-3286-7683]{Tim Lichtenberg}
\affiliation{University of Oxford, Atmospheric, Oceanic and Planetary Physics, Department of Physics, Parks Road, Oxford OX1 3PU, United Kingdom}

\sloppy

%
\begin{abstract}
We demonstrate that \tlnote[the deep volatile storage capacity of] magma oceans \tlnote[has significant] implications for the bulk composition, interior and climate state inferred from exoplanet mass and radius data. \tlnote[Experimental petrology provides] the fundamental properties on the ability of water and melt to mix. So far, these data have been largely neglected for exoplanet \cdnote[mass-radius modeling].
Here, we present an advanced interior model for water-rich rocky exoplanets. The new model allows us to test the effects of rock melting and the redistribution of water between magma ocean and atmosphere on calculated planet radii.
 Models with and without rock melting and water partitioning lead to deviations in planet radius  of  up to 16\% for a fixed bulk composition and planet mass.
This is within current accuracy limits for individual systems and statistically testable on a population level. Unrecognized mantle melting and volatile redistribution in retrievals may thus underestimate the inferred planetary bulk water content by up to one order of magnitude.

\end{abstract}


\section{Introduction}
\sloppy
Many exoplanets discovered to date likely host globally molten mantles -- magma oceans -- because the distance to their stars or their thick atmospheres prohibit efficient cooling and solidification of their rocky mantles \citep{2016SSRv..205..153M,2020SSRv..216...98G}. Magma oceans are substantial reservoirs for volatiles \rev[\citep{2016ApJ...829...63S}], especially water, as water solubility in magma is high compared to other volatile species \citep{2020ChEG...80l5594F}. Above pressures of a few GPa, water even becomes \cdnote[fully] miscible in melt, such that magma oceans can take up significant amounts of water \citep{2016ScChD..59..720N}. Here, we investigate the effect of water in molten rocky mantles on the total radius of the planet and the total volatile abundance that can be inferred from exoplanet observations. \cdnote[We use the term {\it global magma ocean} to highlight that the magma ocean is not locally concentrated or focused on one hemisphere, but its extent may be limited in depth such that deeper parts of the mantle are solid due to increasing pressure in the interior.] 

The volatile abundance of rocky planets alters their structural and dynamical properties \citep{2018haex.bookE..66D}, such as differentiation between core and mantle \citep{2021JGRE..12606724B}, geodynamic regime \citep{2021ApJ...908L..48M}, atmospheric composition \citep{2021SSRv..217...22G}, and thus long-term climate \citep{2020A&A...643A..44S}. The amount of water in particular sensitively controls the rheology of the mantle \citep{2016JGRE..121.1831L} and surface state of the planet, for instance the potential for liquid oceans \citep{2021arXiv210803759M}, and the recycling of life-essential elements between the atmosphere and mantle over geologic time \citep{2016GGG....17.1885F}. The right water abundance to establish habitable conditions on a planet, however, is not a given: too \cdnote[little] of it and the planet remains a desert world; too much of it and the planet turns into a drowned ocean planet. On clement orbits, where water can condense at the surface, more than about $\sim$1 wt\% creates high-pressure ice phases at the mantle-ocean interface \citep{2020SSRv..216....7J}, but even a few 0.1 wt\% exceed the storage capacity of solid planetary mantles \citep{elkins2008linked} and prevent dry land masses on the surface \citep{2018ApJ...864...75K}. Inside the runaway greenhouse threshold surface water evaporates in a feedback loop that drives the planet into a global hothouse climate and melts the surface \citep{1969JAtS...26.1191I}, either directly as a consequence of planetary formation \citep{2013Natur.497..607H}, or due to stellar brightening at a later stage \citep{2013ApJ...765..131K}. The water content of rocky exoplanets is thus an important environmental marker to assess potential habitability and interpret putative biosignatures within the context of a given planetary system \citep{2018AsBio..18..630M,2021AGUA....200294K}.

The orbital transition between clement and runaway greenhouse climates is expected to be abrupt, not continuous, because the outgoing longwave radiation of planets governed by steam atmospheres is controlled by the tropospheric radiation limit \tlnote[at surface temperatures between $\sim$1000--2000 K for approximately Earth-like water ocean inventories in the atmosphere \citep{Boukrouche21}]. Observationally, this predicts that in a system with \tlnote[an initially similar bulk abundance of water with changing] heliocentric distance, planets inside the \rgt should, on average, either (i) be larger than their counterparts outside the \rgt; or (ii) be of similar size but significantly water-depleted. In case (i), planets with similar atmospheric quantities of water in a runaway greenhouse phase are expected to be hot and their atmospheres to be thermally stratified \tlnote[and thus larger], while planets outside the \rgl can condense their water into surface oceans \citep{2019AA...628A..12T,2021Natur.598..276T} and \tlnote[potentially] recycle it into the mantle via tectonic processes. In case (ii), runaway greenhouse climates expose their atmospheric water vapor to high-energy radiation from the host star, which desiccates the planet via \ce{H2O} photolysis and hydrogen loss \citep{2013Natur.497..607H,2013ApJ...778..154W,2015AsBio..15..119L,2016ApJ...829...63S,2019ApJ...881...60A}. The left-over atmospheres, potentially made-up primarily of \ce{CO2} \cdnote[ \citep{ortenzi2020mantle}] or \ce{O2} \citep{2016ApJ...829...63S}, have varying tropospheric radiation limits \citep{2021JGRE..12606711L}, which permits \cdnote[the atmospheres] to cool down and shrink significantly relative to the steam atmosphere case. If mass and radius of an exoplanet are the only two reasonably\cdnote[-well] constrained observational properties, the two end-member cases (i) and (ii) \tlnote[can] lead to order-of-magnitude different interpretation of the maximum water content of planets \cdnote[inside and outside] the \rgt.

However, the equilibrium state of runaway greenhouse climates, where the temperature can increase above the tropospheric radiation limit, lies above the melting point of basaltic rock already for a fraction of the Earth's surface water inventory \citep{Boukrouche21}. Terrestrial and super-Earth exoplanets that start out with significant quantities of nebular H/He can be globally molten for hundred millions to billions of years \citep{KiteBarnett2020PNAS,2021JGRE..12606711L}. Therefore, rocky planets inside the runaway greenhouse state can be expected to be at least partially molten at their surface and in their interiors, if not globally from the primordial magma ocean phase after accretion \citep{2019AA...621A.125B}. Atmospheric volatile compounds can dissolve into liquid or solid rock to various degrees. Water in particular is highly soluble in molten rock, and if a substantial fraction of the planet's interior is molten, most of it will reside in the interior, rather than in the atmosphere \citep[][]{2012AREPS..40..113E,vazan2020new,2021SSRv..217...22G}. 

Retrieving the volatile inventory of rocky planets is a major goal of exoplanetary science to provide quantitative measures of surface habitability \citep{2017ARAA..55..433K,2018AsBio..18..709C} and inform theories on the origin of life on the viability of prebiotic chemical reaction networks on early Earth \citep{2020SciA....6.3419S}. Therefore, we here test the influence of water partitioning and rock melting on the observed radius of planets within the \rgt from transit surveys such as TESS \citep{2015JATIS...1a4003R}, CHEOPS \citep{2021ExA....51..109B}, PLATO \citep{2014ExA....38..249R}, and ARIEL \citep{2021arXiv210404824T,2021ExA...tmp...56H}, \tlnote[as an important prerequisite on the path to inferring the surface state and climate of terrestrial exoplanets with] future direct imaging surveys \citep{HABEX_StudyReport2019,LUVOIR_StudyReport2019,2019arXiv190801316Q,2021arXiv210107500Q}.

\section{Method}
\label{sect:methods}

Our planetary structure model is based on \citet{dorn2015can} and \cite{dorn2017generalized} with significant improvements that are explained in the following. Major improvements are the inclusion of liquid phases both in mantle and core materials and the ability of water to dissolve in magma. The first addition of liquid phases mainly follows the approach taken in Haldemann et al. (in prep.).

\subsection{Interior model}
\label{model}
Our 1-D interior model describes a planet in hydrostatic equilibrium that is composed of three main components: 
an iron-dominated core, a silicate-rich mantle, and a water layer of condensed water or steam, depending on the pressure ($P$) and temperature ($T$) conditions. For the purpose of this study, we do not include an atmosphere of other high mean molecular weight species or H/He, but only account for water. We solve the equations of mass conservation, the equation of hydrostatic equilibrium, thermal transport, and the equation of states for different materials and their phases.
\subsubsection{Metal core}

We assume a core made of Fe and FeS. Many previous studies assume pure iron \citep{zeng2008computational,dorn2015can, rogers2010framework}. However, the addition of light alloys in the Earth's core \citep{dziewonski1981preliminary} can reduce the core density between 5 to 10\%. As data precision increases with ongoing missions like CHEOPS, the addition of light alloys thus becomes important. Possible light alloys include S, Si, O, C, and Ni \citep{2013AREPS..41..657H}. Here, we consider the core to be made of a uniform mixture of Fe and FeS \citep{valencia2007detailed}. To stay consistent with experimental results we use a molar fraction of 10\%, which is well within the maximum molar fraction of 23.4\% that is constrained from laboratory measurements \citep{ichikawa2020ab}.

For pure Fe, we use the equations of state for hexagonal close packed (hcp) solid iron at $P \geq 310$ GPa \citep{hakim2018new} and at $P < 310$ GPa \citep{miozzi2020new}; face-centered cubic (fcc) solid iron  \citep{dorogokupets2017thermodynamics}; and liquid iron at $P \geq 116$ GPa \citep{ichikawa2020ab} and at $P < 116$ GPa \citep{kuwayama2020}. The phase transitions and the melting curve are calculated using \citet{anzellini2013melting}. The presence of sulfur  reduces the melting temperature according to \citep{stixrude2014melting}
\begin{equation}
T_{\rm melt}^{\rm FeS} (P) = T_{\rm melt}^{\rm Fe} (P) \cdot (1 - {\rm ln}{x_{\rm Fe}})^{-1},
\end{equation}
where $x_{\rm Fe}$ is the weight fraction of pure Fe in the mixture of Fe-FeS, i.e., $x_{\rm Fe} + x_{\rm FeS} = 1$. 
For pure FeS, we distinguish between the equations of state for solid FeS \citep{hakim2018new} and liquid FeS  \citep{ichikawa2020ab}. To compute mixtures of Fe and FeS, we use the additive volume law.

The core thermal profile is assumed to be adiabatic throughout the core. At the core-mantle boundary (CMB), there can be a temperature jump as the core can be hotter than the mantle due to the residual heat released during core formation. Following \citet{stixrude2014melting}, this temperature jump depends on the melting temperature of the silicate mantle. If the initially calculated temperature at the CMB is less than the melting temperature of the mantle material, the CMB temperature is increased up to the melting temperature.
\subsubsection{Mantle}

The mantle is made of MgO, SiO$_2$, and FeO that form different minerals. Previous models \citep{dorn2015can} included minor elements like Ca, Al, and Na. Here, the limited availability of high pressure data on their liquid phases lead us to neglect the minor elements that only account for about 7\% of the Earth's mantle mass \citep{workman2005major}. Hence, they introduce little uncertainty to the calculated interiors but add a lot of model complexity.

For the solid mantle, we use the thermodynamical model of \citet{connolly2009geodynamic}, Perple\_X, to compute stable mineralogy for a given composition, pressure and temperature. This model employs equations of state from \citet{stixrude2011thermodynamics}.
For the liquid mantle, we use the equations of state for
Mg$_2$SiO$_4$ from \citet{stewart2020shock}, for
SiO$_2$ at $P\geq 20$ GPa from \citet{faik2018equation}, for SiO$_2$ at $P<20$ GPa from \citet{melosh2007hydrocode}, and for FeO from \citet{ichikawa2020ab}. \ce{Mg2SiO4} was chosen instead of MgO, since \citet{stewart2020shock} recently published an updated version of M-ANEOS with parameters for forsterite, which self-consistently covers a large range in pressure and temperature, which is not available for MgO to our knowledge. To compute mixtures of the above components, we use the additive volume law.

The melting curve is calculated following \citet{belonoshko2005high} ($P<189.75$ GPa) and \citet{stixrude2014melting} ($P \geq 189.75$ GPa):
\begin{equation}
    T_{\rm melt}^{\rm MgSiO_3} (P) =  \begin{cases}
      a_1 \cdot (1+P/a_2)^{a_3} & {\rm if\quad} P< 189.75 {\rm GPa}\\
      b_1 \cdot (P/b_2)^{b_3} & {\rm if\quad} P \geq 189.75 {\rm GPa},
    \end{cases}
\end{equation}
where $a_1 = 1831$ K, $a_2 = 4.6$ GPa, $a_3 = 0.33$, and $b_1 = 5400$ K, $b_2 = 140$ GPa, $b_3 = 0.48$.
The melting temperature is influenced by the composition, specifically the addition of Fe lowers the melting temperature, for which we follow \citet{dorn2018outgassing}:
\begin{equation}
    T_{\rm melt}^{\rm FMS} = \begin{cases}
    T_{\rm melt}^{\rm MgSiO_3} + c_1 \cdot (c_2 - x_{\rm FeO})\\
    T_{\rm melt}^{\rm MgSiO_3} + (c_3 + c_4*P - c_5*P^2) \cdot (c_2 - x_{\rm FeO}),
    \end{cases}
\end{equation}
where $P$ is in GPa, $c_1 = 360$K and $c_2 = 0.0818$ (which is the Earth's iron mantle content \citep{workman2005major}), and other fitting constants are $c_3 = 102$, $c_4 = 64.1$, and $c_5 = 3.62$. \cdnote[Partial melting is neglected.]
\subsubsection{Water layer}
 The water layer is made of pure H$_2$O for which we use the EOS of \citet{haldemann2020aqua}. Depending on pressure and temperature, water can create a steam atmosphere or reside in one of the many condensed ice phases including liquid, ice, and high pressure ice phases.
 
 The transit radius of a planet is assumed to be at a pressure of $P_{\rm Transit}=1$ mbar. This is a simplification as the transit radius depends on temperature, however, the effect on the planets of interest is small \citep{turbet2020revised}. The thermal profile is assumed to be fully adiabatic, except for pressures \cdnote[less than the pressure at the] tropopause \piso, where we keep an isothermal profile. We test two different cases for the tropopause location, i.e., \piso equals 0.1 bar or 1 bar. When assuming \piso equals 0.1 bar, our results are very similar to those of \citet{turbet2020revised} within 0.5\%. If not mentioned otherwise, the temperature at the transit radius is fixed at 400 K to assure water vapour in the atmosphere.

\subsubsection{Water-magma mixtures}
For one of our model scenarios (C, see below), we allow the water to be mixed into the magma. Water naturally dissolves into magma as it thermodynamically equilibrates with silicate melt, leading to hydrous magma oceans. The partitioning between the melt and the steam layer
is determined by a modified Henry's law that accommodates a
power-law relationship between the pressure at the mantle-steam layer boundary (MSB) and the water mass fraction in the melt,
\begin{equation}\label{eq4}
    x_{\rm H2O} = \alpha \cdot (P_{\rm MSB} )^{1/\beta},
\end{equation}
where $\alpha$ and $\beta$ are fitting parameters which are given by \citet{BowerInPrep} (for pressures of 0-300 bar and different rock types) and \citet{2021JGRE..12606711L} (for pressures up to 1GPa). For pressures above 1 GPa, \rev[solubilities reach the so-called second critical endpoint \citep{kessel2005water}. If the second critical endpoint is reached, there is no limit on the amount of water that the magma can take up. Thus, if the second critical endpoint is reached for a given interior model, any addition of water will only be made to the magma reservoir and not to the surface water reservoir. ] We have combined the available fits into an empirical solubility function depending on $P$ (Figure \ref{fig:app}).

Water plays a crucial role in melt composition, with important consequences for the properties of hydrous silicate melt and the melt fraction. Water increases melt fraction by lowering the melting temperature of rock, for which we use the approach of \citet{katz2003new}:
\begin{equation}
    T_{\rm melt}^{\rm FMS, wet} = T_{\rm melt}^{\rm FMS} - d_1*x_{H2O}^{d_2}
\end{equation}
where $d_1 = 43$ K and $d_2 = 0.75$. \cdnote[This relation is limited to the saturation concentration of water in the melt, a regime that we do not enter in our tested cases.]

Water also changes the density of magma and thus modifies the total radius of a planet. However, this effect is  small for low water mass fractions. The melt density decrease per wt\% water depends on the melt composition and is 0.036 g/cm$^3$ for basalts, 0.035 g/cm$^3$ for enstatite and 0.03 g/cm$^3$ for silica. Here, we use a constant value of 0.036 g/cm$^3$ per wt\% water \citep{bajgain2015structure}. According to \citet{bajgain2015structure}, the density contrast between pure melts and hydrous melts is nearly independent of pressure and temperature. At large water mass fractions, a constant density reduction becomes invalid. Thus, we calculate also the density of the rock-water mixture with the additive volume law and use the maximum density of both values. 

In principle, water can also be taken up by solid mantle rocks. However, the solubility of water in solid rocks is orders of magnitudes lower than for magma \citep[below few wt\%,][]{elkins2008linked,2013Natur.497..607H}. Thus, we neglect the effect of hydrated {\it solid} rock on the total radius as it accounts only for a few percent at maximum and only in extreme cases \citep{shah2021internal}. 

\subsection{Interior scenarios}
\begin{figure}[tb!]
 	\centering
 	\includegraphics[width=0.35\textwidth]{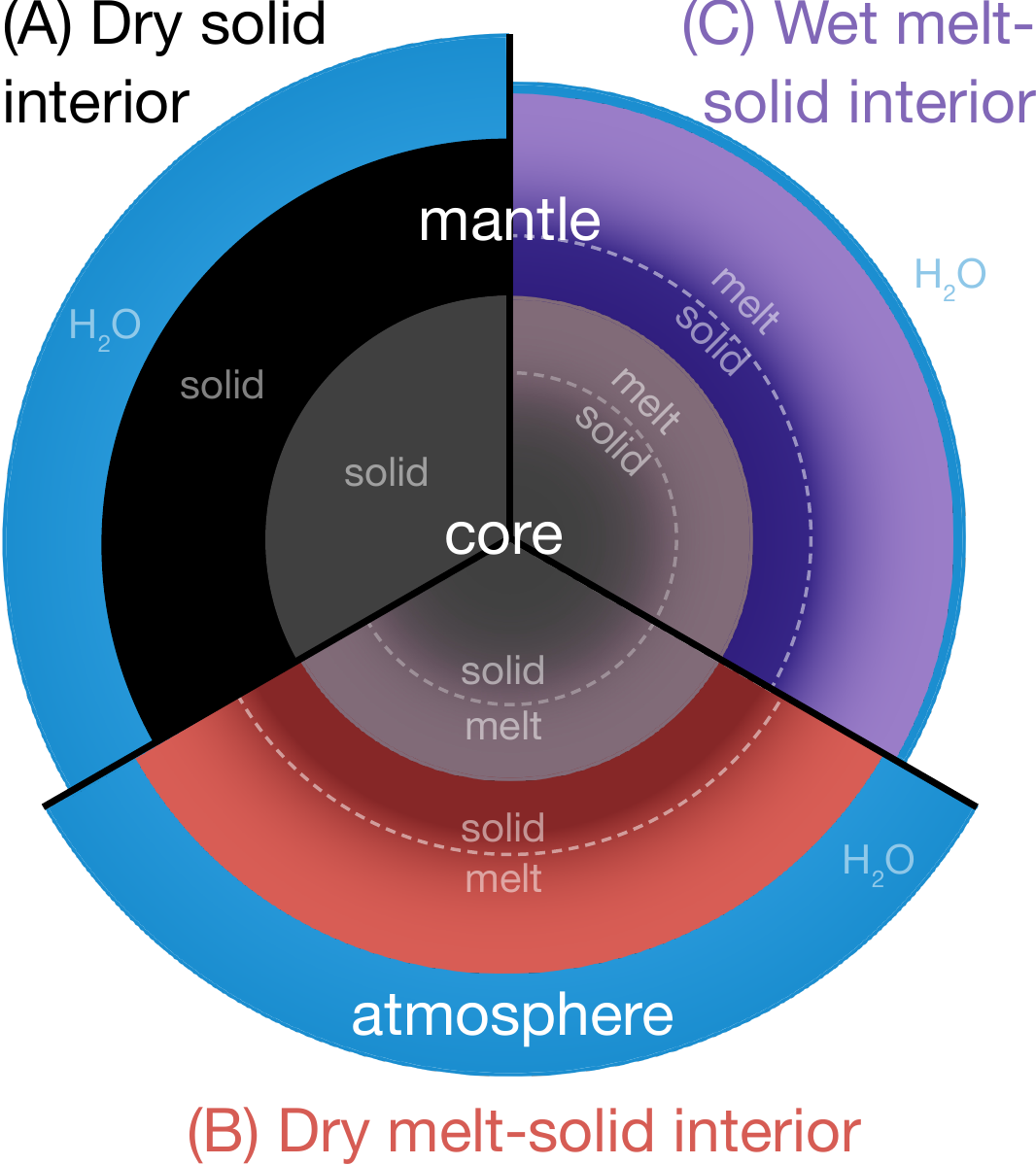}
 	\caption{Three model scenarios employed in our study. Model (A) is very similar to models presented in \citet{dorn2015can} and \citet{dorn2017generalized} and other commonly used exoplanet interior models follow model (A), where liquid rock phases are neglected. Characteristically, the total radius is largest for model (B), where solid and melt phases are present in the core and the mantle. Generally, the radius is smallest for model (C), where additionally the effect of water partitioning into the magma ocean is taken into account. Model C most accurately reflects our current knowledge of mineral physics and exoplanet interiors.}
     \label{fig:1_Scenarios}
 \end{figure}
In order to study the effect of interior models on the amount of inferred water mass, we employ three different interior scenarios:
\begin{enumerate}
    \item[(A)] rocky interior with no melts, and a separate water layer;
    \item[(B)] rocky interior with possible (dry) melt in both mantle and core, and a separate water layer;
    \item[(C)] rocky interior with possible melt in both mantle and core, and water being distributed between mantle melt and a separate surface (steam) water layer depending on the solubility relationship.
\end{enumerate}
Figure \ref{fig:1_Scenarios} illustrates all model scenarios, demonstrating that model (B) generally features the largest radii for a given planetary mass, while model (C) generally features the smallest radii. All interior scenarios follow the model described in section \ref{model} with few exceptions for (A) and (B).
In interior scenario (A), we artificially set the adiabatic temperature profile in the rocky interior to low temperatures, i.e., the temperature at the water-mantle boundary is fixed to a maximum of 1500 K. This prevents the occurrence of melts in the planets of interest and thus water can only be present in a separate layer on top of the mantle. In interior scenario (B), water solubilities are fixed at zero.

\section{Results}
\label{Results}

\begin{figure}[tb!]
	\centering
	\includegraphics[width=0.45\textwidth]{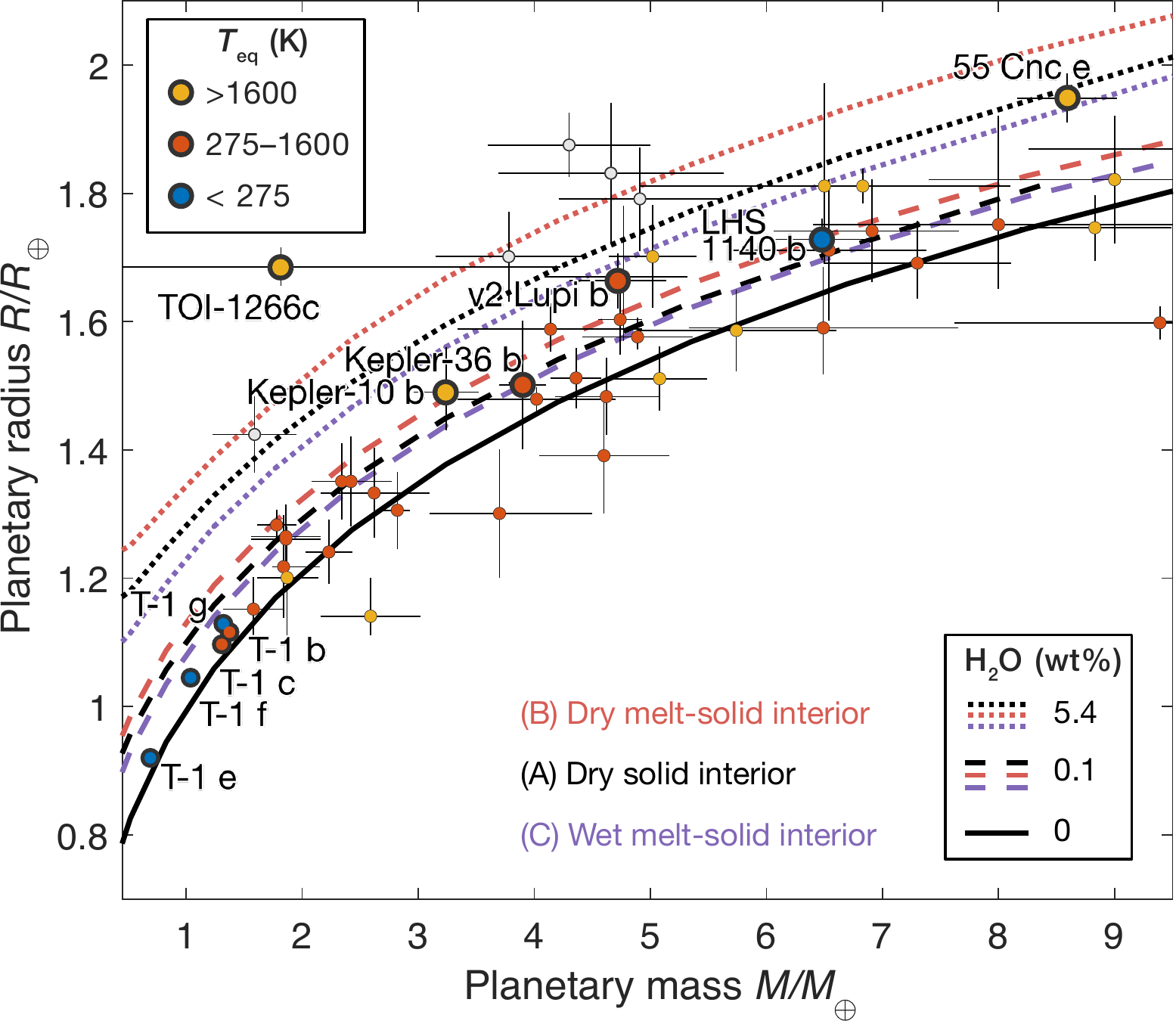}
	\caption{Mass-radius diagram for exoplanets from the PlanetS database \citep{otegi2020revisited} (as of Oct 2021), TOI-1266c (see discussion) and curves for different composition and interior models: the rocky interior is composed of Mg-Si oxides and silicates (66\% in mass) and an iron core of 33\% in mass. In addition, varying mass fractions of \cdnote[bulk] water are added: 5.4 wt\% (dotted lines), 0.1 wt\% (dashed lines), or = 0 wt\% (solid black line). The black (A), red (B), and purple (C) coloring of the lines refers to the different model scenarios in Figure \ref{fig:1_Scenarios}. Largest radii are achieved when accounting for magma oceans while neglecting the solubility of water in magma (scenario B). Taking water solubility into account reduces the calculated radii by $\sim$5--15\%. The transit radius on this figure is fixed at 400 K and 1 mbar. Individual planets are colored according to their equilibrium temperature.}
    \label{fig:2_MR}
\end{figure}

We explore the effect of water in the model scenarios A, B, and C on exoplanet transit radii. The calculated mass-radius curves in Figure \ref{fig:2_MR} illustrate that transit radii differ significantly for a given planetary mass and water mass fraction. These deviations are on the order of commonly achieved observational uncertainties for individual planets. In general, calculated radii are largest if a planet's interior accounts for liquid rock phases (internal magma oceans) and all water is surface water (scenario B). Imposing the assumption that all mantle rocks are in solid phase reduces the calculated radii by up to several percent (scenario A). A further critical reduction in radius is obtained if water is distributed between the surface and the magma ocean as determined by its equilibrium stage (scenario C). 

The radius reduction of hydrous magma ocean planets (C) compared to magma oceans with only surface water (B) of identical mass fraction is because water dissolved in the magma ocean contributes less to the total radius compared to the same mass in the surface/atmospheric reservoir. On the atomic scale, water is a relatively small molecule that easily fits inbetween the molecules of silicate melts and thus the increase in volume of hydrous magma compared to dry magma is  limited.

\begin{figure}[tbh]
	\centering
	\includegraphics[width=.45\textwidth]{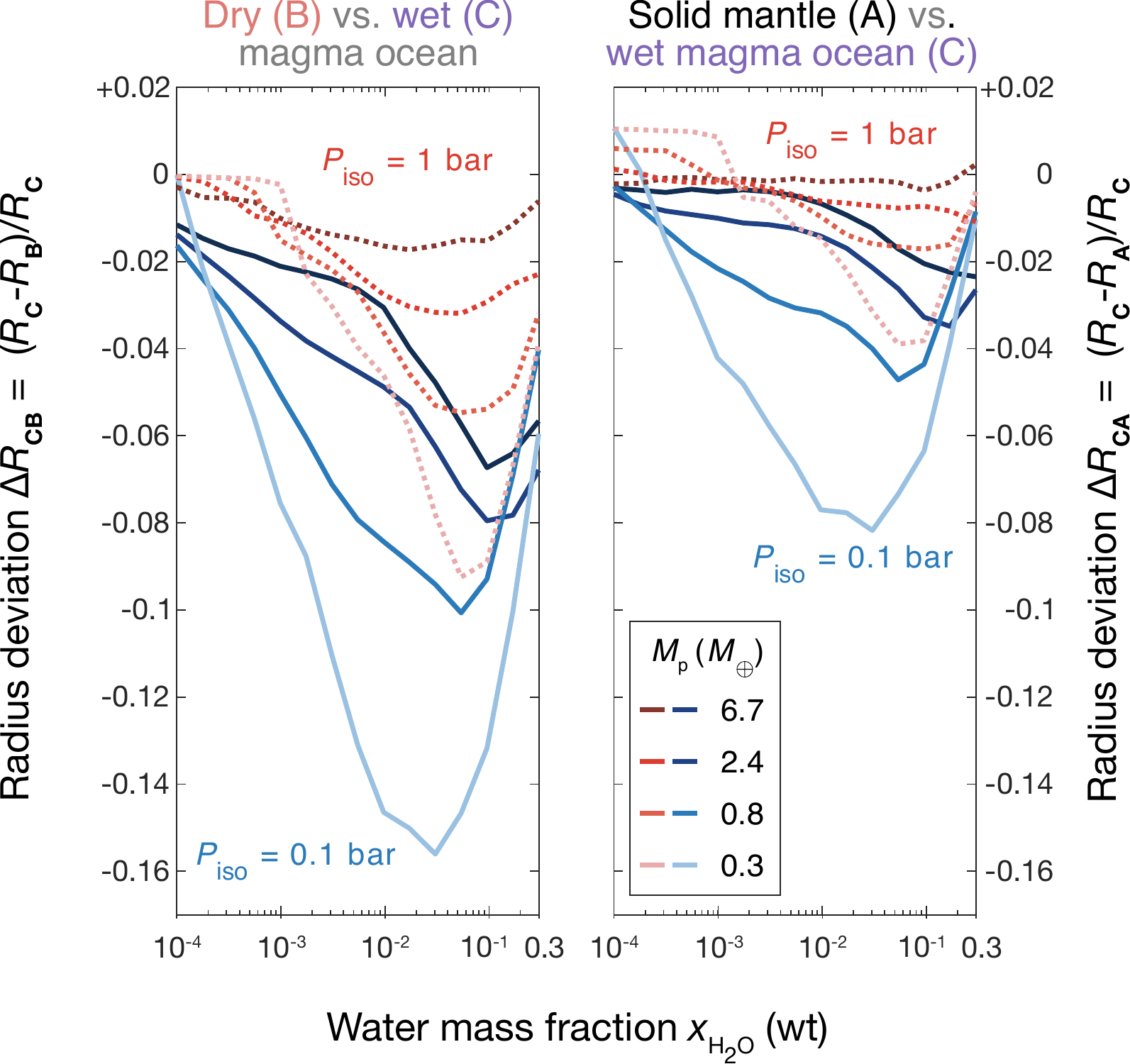}
	\caption{Calculated planet radii depending on the interior model scenarios (A, B, C) and as a function of \cdnote[bulk] water mass fraction. {\it Left}: Comparing scenario C to B indicates that errors of up to 16\% are produced when correctly accounting for liquid rock phases but neglecting water solubility. {\it Right}: Comparing scenario C with the most commonly used model in the exoplanet literature, scenario A. When neglecting both liquid rock phases and water solubilities, the error on the radius is 1--8\% for the cases shown. Largest errors are identified for small planet masses. The transit radius in all scenarios is fixed at 400 K and 1 mbar.}
    \label{fig:3_deltaR}
\end{figure}

The calculated radius deviations between the three model scenarios (A, B, C) depend on bulk planetary mass, water mass fraction, and the melt fraction of the silicate mantle. Figure \ref{fig:3_deltaR} quantifies relative radii differences between the three scenarios. We find increasing differences of up to 16\% for decreasing planetary masses and between 1--10 wt\% water mass fractions, when comparing scenario B and C (left panel). Increasing the planetary mass reduces the differences in transit radius. However for a 6.7 \ME planet, differences in radius are still significant with a deviation of 7\% for $P_\mathrm{iso} = 0.1$ bar. When comparing model scenarios A and C, radius differences are a factor $\sim$ 2 smaller. This is because different simplifications imposed in scenario A partly cancel each other out.  More specifically, scenario A neglects the effect of liquid rock, which {\it decreases} the radii, and it neglects the effect of water dissolution in the magma underneath, which {\it increases} the radii. Both simplifications compensate each other which is why the radii differences between model A and C are smaller than between B and C.

In Figure \ref{fig:3_deltaR} we also test the effect of the thermal profile in the steam atmosphere. We test a potentially varying location of the tropopause by changing the pressure \piso, below which we assume an isothermal profile. Steam atmospheres with \piso of 0.1 bar are thus overall warmer than those with \piso of 1 bar. There are two consequences on the calculated interior profiles: Firstly, warmer atmospheres lead to larger atmospheric scale heights and thus larger total radii (see Section \ref{sec:_trappist} for examples). Secondly, warmer atmospheres allow a larger portion of the mantle underneath to be molten and thus more water can be dissolved in the magma. For the relative radii differences in Figure \ref{fig:3_deltaR}, the latter effect can be clearly seen as radii differences of cooler atmospheres (with \piso of 1 bar, dotted lines) are a factor of $\sim$ 2 smaller than the differences among warmer atmospheres (with \piso of 0.1 bar, solid lines).

Figure \ref{fig:3_deltaR} shows a characteristic behaviour: for small water mass fractions below 1 wt\%, the relative differences between the radii decrease \tlnote[from the peak between $\sim$3--10 wt\% water]. \tlnote[Similarly,] for large water mass fractions of tens of percent, the relative differences decrease from the peak differences, although they never reach zero.
This is because for small water mass fractions, an increase in \wmf leads to higher surface temperature at the top of the magma ocean and thus the melt fraction of the mantle increases and thus more water can be dissolved in the mantle. In consequence the radius difference increases as more water from the surface water reservoir can be stored in the magma ocean mantle. 

At large water mass fractions (tens of percent), the melt fraction of the mantle becomes 100\%, the mantle is completely molten and therefore a further increase in relative radius deviation between model scenarios is not possible. In this stage gravitational compression becomes the dominant effect. In essence, the radius does not increase linearly with surface water fraction. For example, the radius difference between planets of 0 and 5 wt\% steam envelope is $\sim$3 times larger than between planets of 5 and 10 wt\% steam envelope. Thus for large \wmf, different {\it surface} water budgets only yield small radius differences. The increasing values of relative radii differences for large \wmf in Figure \ref{fig:3_deltaR} have reached this regime.

In other words, the differences in radii are caused by (1) different mantle melt fractions and their ability to store large amounts of water as well as (2) the surface water budgets. Wherever relativ radii differences decrease with increasing water mass fractions in Figure \ref{fig:3_deltaR}, the first effect dominates (small \wmf), while the second effect dominates where relative radii differences increase (large \wmf).

\begin{figure}[tbh]
	\centering
	\includegraphics[width=.45\textwidth]{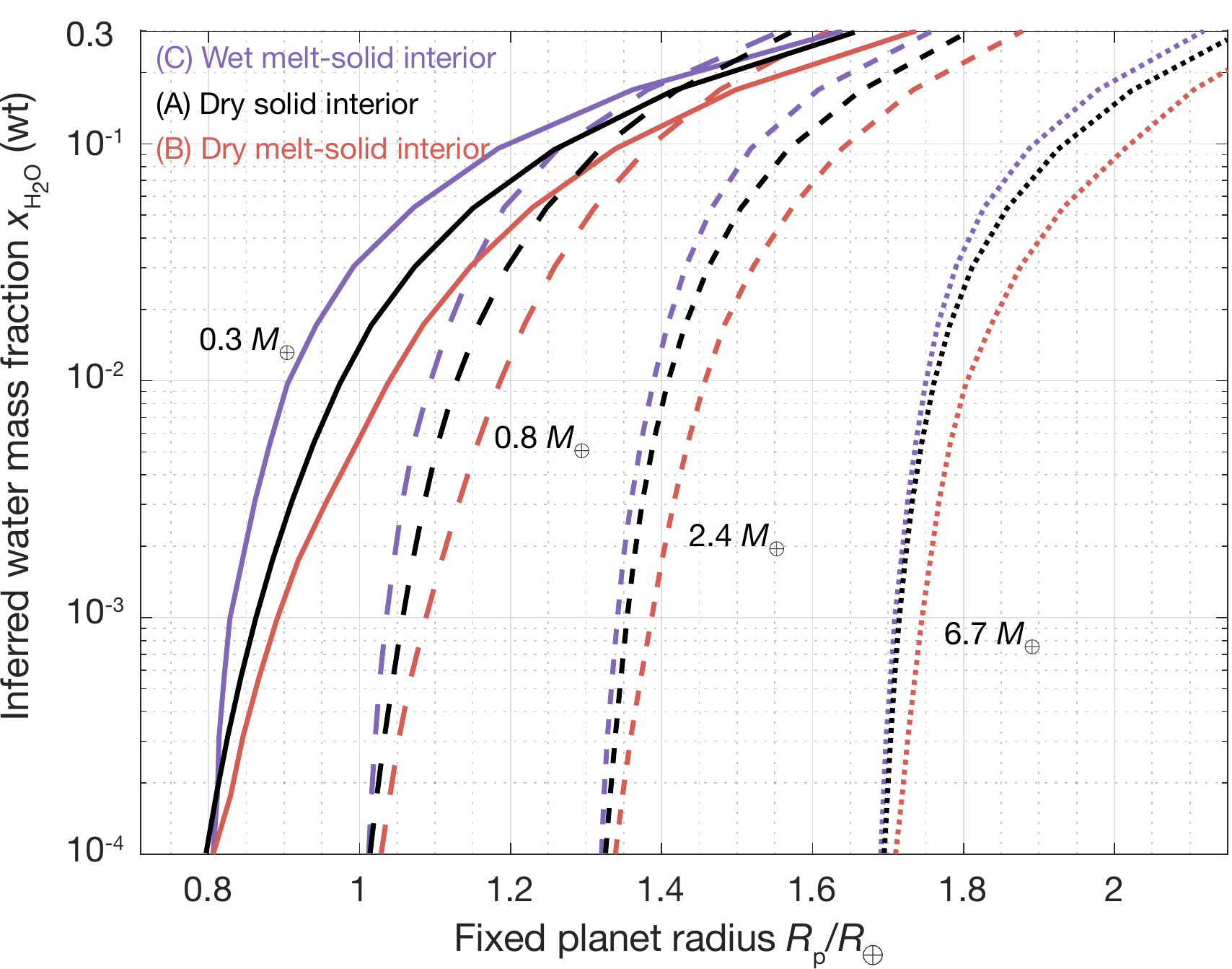}
	\caption{Inferred water mass fraction depending on interior model scenarios (A, B,C ) and as a function of planet transit radius. The inferred \cdnote[bulk] water mass fraction can vary by up to one order of magnitude depending on the interior scenario used. This is clear from the vertical difference between different colors for planets of the same mass (different line styles). Largest differences between the scenarios are seen when comparing scenario B (red) to C (purple), while differences between scenarios A (black) and C are a factor 5 at most. If degeneracies of rocky interior composition were added, the effect on inferred water mass fraction would be similar. \cdnote[The curves for 0.3 and 0.8 \ME cross for high water mass fractions, as steam atmosphere become gravitationally unbounded for low planet masses.] Again, the transit radius in all scenarios is fixed at 400 K and 1 mbar.}
    \label{fig:4_deltawmf}
\end{figure}

As these results indicate, the calculated radii sensitively depend on the interior model scenario used. In consequence, for a given planetary mass and radius from exoplanet observations of individual planets, the inferred water mass fraction will depend on the interior model. Figure \ref{fig:4_deltawmf} illustrates how the inferred water mass fraction for a fixed (observed) exoplanet radius differs for the model scenarios (A, B, C). The differences between water mass fractions between scenarios B (lowest \wmf) and C (highest \wmf) are on the order of one magnitude. The differences are a factor of two smaller when comparing scenarios A and C. Figure \ref{fig:4_deltawmf} demonstrates that the phase state of the planetary interior affects the estimated water mass fraction of an exoplanet significantly. Hence, estimated water mass fractions obtained using different interior models are not easily comparable. 
In Section \ref{discussion}, we discuss selected planets and their estimated water mass budgets.

\section{Discussion}
\label{discussion}

A large portion of detected super-Earths may be magma ocean worlds. Out of the 54 exoplanets shown in Figure \ref{fig:2_MR}, 12 (22\%) of them have equilibrium temperatures above 1600 K which is already hot enough to melt surface rocks, independent of any atmosphere. Furthermore, 47 planets (87\%) have equilibrium temperatures above 400 K. If these planets host even small amounts of greenhouse gases \citep[a few tens of bars in case of \ce{H2O} or \ce{H2},][]{KiteBarnett2020PNAS,2021JGRE..12606711L,Boukrouche21}, their surface rocks can be molten, allowing a deep magma ocean underneath. At a fraction of one Earth ocean (EO) \citep[Earth hosts about $\sim$3--11 oceans in its mantle+atmosphere and up to 100 oceans locked in H in the core,][]{PeslierISSI2018} irradiation above the runaway greenhouse limit ($\sim$280 W m$^{-2}$) will drive the planet into the globally molten phase. For instance, at 50 bar surface pressure of water, the surface equilibrium temperature would be 1500 K, at 260 bar ($\sim$1 EO) the surface temperature would be $\sim$1800 K \citep{Boukrouche21}. From an astronomical perspective, atmophile compounds are abundant in star-forming and thus planet-forming regions \citep{obergbergin2021}. Migration of planets, icy pebbles, and snowlines suggest that inner planetary systems are easily enriched in major volatile compounds \citep{2019NatAs...3..307L,2019AA...624A.109B}. The water snowline is the preferred nucleation region for the growth of planets \citep{2017A&A...608A..92D}, including the terrestrial planets of the Solar System \citep{2021Sci...371..365L}. Even in the absence of exogenous volatiles, super-Earth mantles may generate H$_2$O by internal geochemical reactions \citep{2021ApJ...909L..22K}. Therefore, super-Earths with outgassed secondary atmospheres may preferentially harbour magma oceans unless they are bare rocky worlds. Although the absence of any atmosphere is very difficult to achieve, relatively thick atmospheres can be ruled out in some cases. For the case of LHS~3844b, atmospheres with surface pressures above 10 bars could be ruled out \citep{kreidberg2019absence}; and indeed a 10 bar atmosphere of steam would not be sufficient to raise the surface temperature high enough for rocks to be molten on LHS~3844b.

If many super-Earths can in principle be magma ocean worlds, they may host consequentially large volatile reservoirs in their interiors. So far, these deep mantle reservoirs are not taken into account when inferring volatile (or water) budgets from planetary data. Previous studies have published inferred  {\it surface} water contents \citep{rogers2010framework, dorn2017bayesian,unterborn2018updated,turbet2020revised, brugger2017constraints,2019ARA&A..57..617M,2020ApJ...896L..22M,2021ApJ...914...84A} or are limited by only accounting for {\it surface} water reservoirs when modelling interiors \citep{venturini2020nature, owen2017evaporation,fortney2013framework,2021MNRAS.503.1526R}. Our study highlights that there is a clear difference in {\it bulk} water contents versus {\it surface} water content. Taking magma ocean reservoirs into account can increase water budget estimates by one order of magnitude for a given radius or it can reduce calculated planetary radii by up to 16\%. This is equivalent to a change in planetary density of 75\%. Clearly,  there is a need to reinvestigate possible water budgets for the populations of observed super-Earths and sub-Neptunes.

Here, we focus on water in magma, as solubilities of other volatiles (e.g., CO$_2$, H$_2$, CO, CH$_4$) are  1--6 orders of magnitudes lower than for water and partial pressures below 1 GPa. Thus, unless large quantities are available of other species, water will be the dominant volatile in the magma.

\ce{H2} can in fact be available in large quantities when a planet has accreted primordial H/He envelopes.  Hydrogen-dominated envelopes are likely shaping the radii of sub-Neptunes \citep{owen2017evaporation,fortney2013framework}, although additional large water budgets are possible \citep{mousis2021,venturini2020nature, kite2020atmosphere}. But what about deep water reservoirs in molten mantles of sub-Neptunes? 

Primordial hydrogen may reduce mantle oxides and produce H$_2$O \citep[potentially dominant when Fe reaches the core,][]{2021ApJ...909L..22K} or FeH, but it is unclear if reduced Fe would merge with the core \citep[][]{2021ApJ...914L...4L}. In global chemical equilibrium, mantle oxygen may be able to produce water in mole number comparable to or more than hydrogen \citep{schlichting2021chemical}. Thus, primordial H/He atmospheres may be enriched with water vapour \citep{kimura2020formation}. In consequence, the water will in large parts dissolve into the mantle. Thereby, \cdnote[atmospheric water abundances of sub-Neptune envelopes would consequently become sub-solar]. For now, there is no comprehensive interior model that accurately accounts for the chemical reactive atmosphere-magma ocean boundary as well as the partitioning of water and other volatiles in the deep interior of sub-Neptunes.  Only few studies have investigated individual aspects of it \citep{2020ApJ...891..111K,2021ApJ...914L...4L,schlichting2021chemical,vazan2020new,kimura2020formation,olson2018hydrogen,chachan2018role}. As volatile partitioning in the deep interior has been neglected for inference studies of exoplanets, previous interior predictions have thus far generally underestimated the amount of water and hydrogen for sub-Neptunes. \cdnote[Sub-Neptune envelopes may possess compositional gradients \citep{ormel2021planets,helled2017fuzziness}, e.g., water might only be mixed within a hydrogen layer up to heights where water condenses. This effect itself influences the calculated radii and should ideally be considered in parallel with the partitioning of volatiles in the deeper planetary parts.] This is particularly important for very water-rich atmospheres as a potential explanation for the radius valley \citep{2020ApJ...896L..22M,2021ApJ...914...84A}: most of the atmospheric water will be dissolved in the deep interior and does not contribute to the atmospheric layer thickness.

Our presented model is a static model and does not focus on evolutionary aspects. Atmospheric escape can efficiently reduce the amount of surface water by high energy stellar flux \cdnote[\citep{johnstone2020hydrodynamic}]. A reduction of surface water mass will shift the equilibrium state between magma ocean and surface reservoir according to Eq. \ref{eq4}, such that fractions of dissolved water outgas and are added to the steam atmosphere.
A planet with water in its magma ocean (scenario C) will experience less water-loss compared to a planet with only surface water (scenario A and B). This is because the partitioning of water in the interior \tlnote[further decreases the upper-atmosphere water mixing ratio due to condensation \citep{2021arXiv210812902G}]. Hence, high solubility of H$_2$O in magma oceans may enable its safe storage over long time spans. Time-dependent coupled models of magma ocean evolution, outgassing and atmospheric escape are necessary to tackle these questions \citep{KiteBarnett2020PNAS,2021JGRE..12606711L}.

Here, we have not addressed the inference of interior parameters from specific exoplanet data but have outlined the physical reasoning and implications of magma ocean interiors on transit radii and physical state of the planet. This interior model introduces a higher degree of non-linearity compared to other interior models \citep{unterborn2016scaling, dorn2017generalized,2020ApJ...896L..22M}, as there are stronger inter-dependencies between interior parameters. It does not, however, introduce a higher degree of degeneracy. Thus using our model in an inference scheme does not lead to larger uncertainties on predicted interior properties. The new model comes with a higher computational cost: it is two times slower than the previous models from \citet{dorn2017generalized}. However,  inference schemes can be done more efficiently\footnote{Some model parameters can be computed extremely fast (e.g., planet mass), while others require a detailed structure computation (e.g., planet radius). For a more efficient inference, a detailed structure computation is only done, if a newly proposed set of model parameters is {\it not} rejected, assuming a perfect fit for model parameter that would require an expensive computation.} such that the overall computational cost for an inference analysis that includes our new model is two times {\it faster} than previously \citep{dorn2017generalized}.

In the following we consider the potential influence of magma oceans on specific exoplanets.

\subsection{Trappist-1 planets}
\label{sec:_trappist}

The Trappist-1 system \citep{gillon2017seven} hosts a fascinating chain of small and cool planets. Our current knowledge on their masses and radii \citep{agol2021refining} allows to gain an improved understanding of their interiors and atmospheres. Possible scenarios that fit the slightly lower densities compared to an Earth-like interior include: (1) rocky interiors that are iron-depleted compared to the Earth, (2) a core-free rocky interior with oxidised iron, (3) surface water reservoirs on top of a rocky interior, or possibly (4) H-rich cores \citep{schlichting2021chemical}.

Interestingly, single \cdnote[composition mass-radius] curves exist that can fit all seven planets \citep{agol2021refining}. This is possible for the purely rocky, iron-depleted scenarios (1 \& 2). A single interior model that can explain them all, seems to be intriguing as the least complex solution.  

For the water-rich scenario (3), the amount of water that is compatible with each individual planet strongly depends on their individual equilibrium temperatures. While the inner planets b, c, and d seem to feature small water mass fractions ($< 10^{-5}$) in the form of steam atmospheres, the outer planets can host up to several wt\% of condensed water \citep{agol2021refining}. Thus, if the lower densities of the Trappist-1 planets are due to a higher amount of water compared to the Earth \citep{2019A&A...627A.149S,dorn2018interior,2019NatAs...3..307L}, there is no single interior model that can explain all planets. 

\begin{figure}[tbh]
	\centering
	\includegraphics[width=0.45\textwidth]{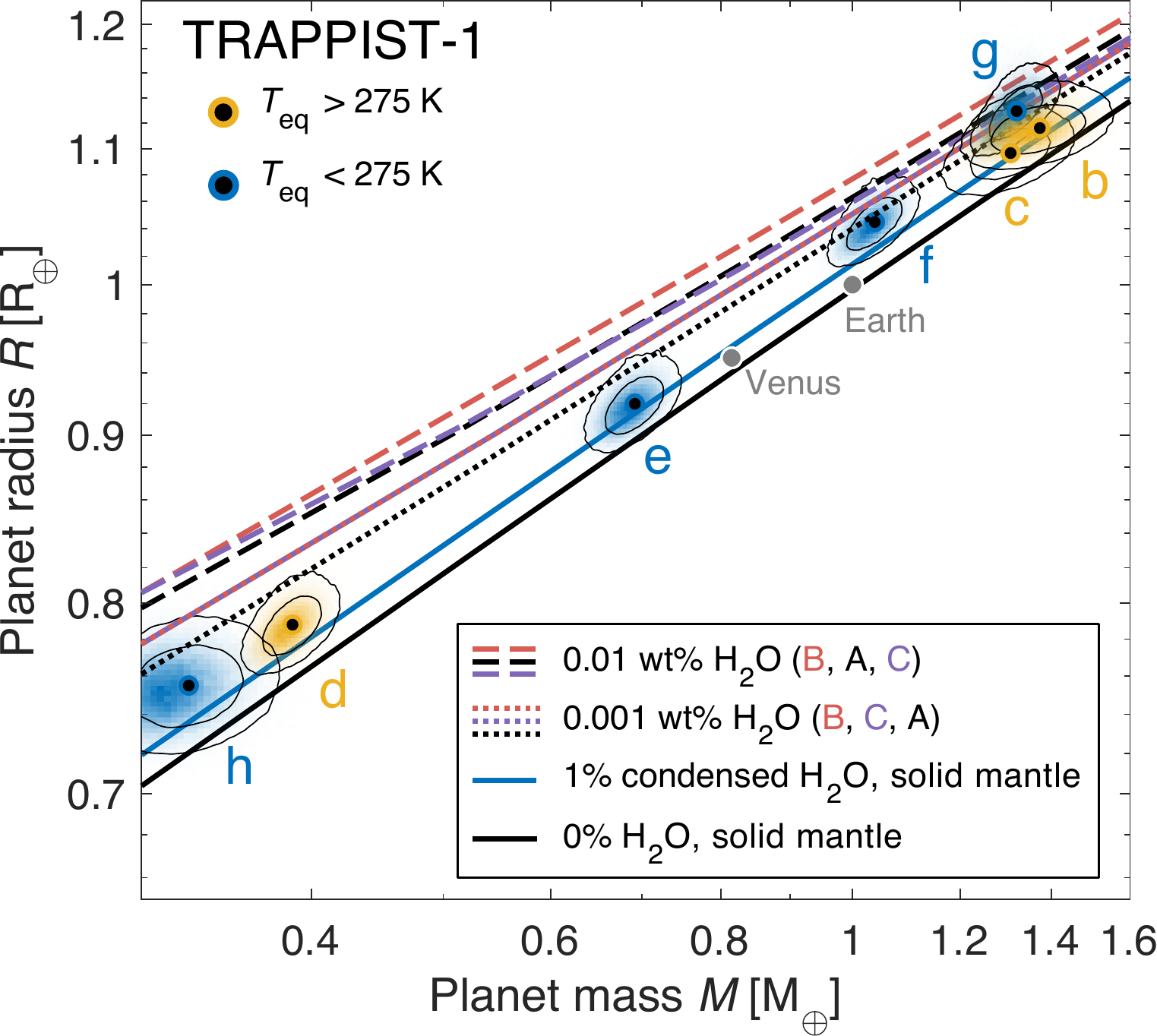}
	\caption{Mass-radius diagram showing the seven confirmed Trappist-1 planets \citep{agol2021refining} and curves for different interior models: the rocky interior is composed of Mg-Si oxides and silicates (66\% in mass) and an iron core of 33\% in mass. In addition 0.01 \% mass fraction of \cdnote[bulk] water are added. \cdnote[The letters refer to rocky interiors being (A) solid \& dry, (B) possibly molten \& dry, (C) possibly molten \& wet, see] Figure \ref{fig:1_Scenarios}. For the outer planets (e-h), steam atmospheres do not apply, instead water is condensed on the surface (blue curve). A magma ocean with dissolved water only establishes for \piso = 0.1 bar and high masses ($>$0.5\ME). For \piso = 0.1 bar and masses below 0.5 \ME, no magma ocean is present and thus model scenario (C) provides identical results compared to scenario (B). Dotted lines show interiors for which we artifically imposed a mantle surface temperature of 1800 K (B \& C).}
    \label{fig:2_MR_Trappist}
\end{figure}

The timing of magma ocean crystallization sensitively depends on the initial amount of volatiles inherited from planetary formation. Hence, this begs the question whether the inner planets can have a magma ocean water reservoir at present day. Is it possible that all planets actually started off with similar water contents while the inner planets have most of their water lost to space or stored in molten mantles? Figure \ref{fig:2_MR_Trappist} shows that within 2-$\sigma$ uncertainty, planets b and c are consistent with 0.01 wt\% of water at maximum (dotted purple line). For water mass fractions below 0.01\%, no hydrous magma ocean is present. This means that for lower water mass fractions, the steam atmosphere does not allow to raise the temperature at the surface high enough to keep mantle rocks molten, or $P_{\rm MSB}$ is too low and dissolved water budgets are marginal. In these cases, the interior model (C) yields identical results compared to model (B) (masses below $\sim$0.4 \ME in Figure \ref{fig:2_MR_Trappist}). Thus, if no other interior heating mechanism is considered, it is unlikely that the inner planets posses magma oceans at present day in which large quantities of water could be stored. This is consistent with the planets being depleted in atmospheric water by \ce{H2O} photolysis and atmospheric hydrogen escape.

Indeed, heating mechanisms are actually proposed to maintain magma oceans on the inner planets  \citep{bolmont2020impact,kislyakova2017magma,barth2021magma}.  However, even if we force the interior model (C) to allow for magma oceans  by fixing the temperature of the mantle surface to 1800 K (Figure \ref{fig:2_MR_Trappist}, dotted lines), the amount of dissolved water is marginal. In this case, the curves of scenarios (C) and (B) appear indistinguishable. Hence, if the inner planets have magma oceans, the amount of dissolved water must be limited to minor amounts.

\subsection{\lupi}

Previous models for \lupi estimate a total water amount of $12.6^{+14.7}_{-11}$ wt\% and a low core mass fraction of 14\% \citep{delrez2021transit}. Such large amounts of water can only be achieved if water is assumed to be fully condensed. Given the equilibrium temperature of 905 K, this is unlikely to be the case. The planetary density is significantly below an Earth-like interior (see Figure \ref{fig:2_MR}). The low density can be explained by a molten mantle and the presence of water. 

Water will form a steam atmosphere and be in large portions hidden in the underlying magma ocean.
More specifically, the planet can be described with a steam atmosphere of about 0.07 wt\%, while the mantle contains 2 wt\% of water in the melt (melt fraction of 40\%). For this case, we assume an Earth-like core mass fraction. A smaller core mass fraction similar to \citet{delrez2021transit} would further decrease the inferred total water budget. Our estimates thus differ significantly from those of \citet{delrez2021transit}, which can be in part attributed to the different interior models but also to the treatment (or the inclusion) of a steam atmosphere. A complete inference analysis within the context of our model is beyond the scope of this paper.

\subsection{55~Cnc~e}

The intensely irradiated super-Earth 55~Cnc~e features an unusually low density \citep[$\rho=1.164\pm0.062 \rho_{\oplus}$ at a mass of $M_{p}=8.59\pm0.43$\ME,][]{crida2018mass,crida2018massb}. Its interior has previously been interpreted to be due to a hydrogen-rich or nitrogen-rich volatile layer \citep{modirrousta2020hot,hammond2017linking}, an absent or a thin metal-rich atmosphere \citep{ridden2016search,morris2021cheops} on top of a magma ocean \citep{2019AA...631A.103B}, or an interior enriched in those minerals (e.g., Al, Ca) condensing at high temperatures and thus depleted in iron \citep{dorn2018new}. Observations in Ly-$\alpha$ indicate the absence of a primordial atmosphere, while the presence of water could not be excluded although its presence was evaluated to be unlikely \citep{bourrier201855}.

Any atmosphere for this hot planet ($T_{\rm eq} \approx 2000$ K) is in chemical equilibrium with the rocky and necessarily molten interior. Volatiles like water could be contained in the magma ocean. Previous estimates on water loss were limited to \emph{surface} water reservoirs \citep{bourrier201855} and did not account for deep reservoirs where water is shielded from XUV irradiation. We thus hypothesize that water could still be present in the interior of 55~Cnc~e, mainly in the deep mantle reservoir. If this were the case, we expect non-zero partial pressures for water in a metal-rich atmosphere, or more specifically ionized water \cdnote[given its stellar irradiation environment]. Hence, a plausible atmosphere could contain H, O, Mg, Si, S, as well as other minor metal elements. A more tailored escape model would be needed to estimate the possible loss of hydrogen in light of deep magma ocean reservoirs for water.

Given the data of 55~Cnc~e, it is possible that the planet has several percents of water in the melt. For a melt fraction of 50\% in the mantle, this would account for 1 wt\% of dissolved water relative to the total mass of the planet. In that case, the surface reservoir would only contain less than 0.01 wt\% of water. Such an interior can explain the low density of 55~Cnc~e while only little water is exposed on the surface to evaporative loss. It remains to be seen whether 55~Cnc~e must have lost all its water during its lifetime or if deep water reservoirs in the molten interior can prolong the retention of it.

\subsection{Kepler-10b and Kepler-36b}

Both Kepler-10b and Kepler-36b have high equilibrium temperatures (2169 K and 979 K, respectively) and both planets are less dense than Earth (Figure \ref{fig:2_MR}). Hence water is a possible component to explain their bulk densities.

For Kepler-10b, a purely rocky interior of 3.3 \ME in mass that fits the stellar abundance proxy \citep{dorn2015can} with [Fe/H] = -0.14, [Mg/H] = -0.01, and [Si/H] = -0.1, yields a radius of 1.429 \RE, which is 3\% lower than the median observed radius of $1.47\pm0.03$ \RE \citep{adibekyan2021chemical}. The addition of a thin steam atmosphere of $1.7 \times 10^{-6}$ wt\% allows to match the observed radius. In that case, $\sim$0.01 wt\% of the planet's mass is contained as dissolved water in the mantle. 
Our estimates for the {\it surface} water reservoir are similar to previous estimates \citep{dorn2017bayesian}, which employed a model that is close to model scenario (A) with a dry solid-only rocky interior. Of course, other volatiles could dominate the atmosphere (e.g., CO$_2$), which would imply their presence in the magma by necessity.

For Kepler-36b, a pure rocky interior of the planet's mass of 3.9 \ME \citep{otegi2020revisited}
yields radii of 1.44 to 1.48 \RE (for core mass fractions between 0.35 and 0.25, respectively) which is slightly below the observed radius of $1.5\pm 0.1$ \RE. The addition of $1 \times 10^{-4}$ wt\% surface water would allow to fit the radius (for a core mass fraction of 0.33), and it would imply an additional water reservoir in the molten mantle of $\sim$0.01 wt\% of the total mass. The presence of water for Kepler-36b is further supported by its proposed formation path \citep{owen2016initial}, where it would have formed with a thick H/He envelope. Any primordial envelope is chemically reducing and would lead to the formation of water \citep{kimura2020formation}.

\subsection{TOI-1266c} 

The planet candidate TOI-1266c was recently hypothesized to host a steam atmosphere \citep{harman2021snowball}. Although TOI-1266c has a large uncertainty on mass and thus can be fit with very different interiors, it is likely a volatile-rich planet given its peculiar low density (Figure \ref{fig:2_MR}). Indeed, the nominal data of TOI-1266c \citep[1.9 \ME, 1.673 \RE,][]{stefansson2020mini} allows for fully molten mantle with a total water budget of 31 wt\% (for a core mass fraction of 33\%). In this case, the total water budget of 31 wt\% is divided between the surface reservoir (5 wt\%) and the magma ocean reservoir (26 wt\%). Our proposed interior composition is significantly different from those models with 50--77\% water proposed by \citet{harman2021snowball}, who refer to different mass-radius-relationships \citep{fortney2007planetary,noack2016water,zeng2019growth}. The large differences stems from the fact that these pre-calculated mass-radius relationships do not account for liquid phases, nor the dissolution of water in melt, nor the specific equilibrium temperature of TOI-1266c. Hence, these mass-radius-relationships are unable to provide self-consistent interiors for the conditions of TOI-1266c.

\section{Summary and Conclusions}
\label{Conclusions}

The majority of observed super-Earths and sub-Neptunes are subject to irradiation that promotes molten rocky interiors, i.e. magma oceans, either due to direct surface melting or greenhouse forcing from water or other volatiles. Global magma oceans represent potentially vast volatile reservoirs that can lock them in the deep interior of planets and consequently shrink the atmospheric reservoir mass. Here, we focused on water and its ability to be stored in large quantities in magma oceans due to its high \cdnote[solubility] in molten rock relative to other volatile compounds. In contrast to previous work \rev[\citep{2016ApJ...829...63S,2019AA...631A.103B,2021JGRE..12606711L}], we account for the entire range of low water solubilities up to the miscibility regime and calculate the effect on mass-radius relationships by providing a comprehensive interior model.

We demonstrate that magma ocean planets with volatiles locked up in the interior differ fundamentally from their solid or molten mantle counterparts with all volatiles stored in the atmospheric reservoir. The differences concern not only interior structure, but also likely their evolution.
\cdnote[Our work complements previous studies that suggest water-dominated secondary atmospheres to emerge for rocky planets evolving from sub-Neptunes with primordial \ce{H2}-dominated atmospheres \citep{2021ApJ...909L..22K}.]

Introducing a quantitative treatment of volatile storage in molten rock phases is essential for correctly inferring bulk \cdnote[volatile mass budgets], and thus being able to retrieve the formation and volatile loss chronology of specific exoplanets and their environment. This will be essential to probe the atmospheric and climate diversity of rocky exoplanets, and to be able to use short-period exoplanet abundances as environmental proxies for their more distant, potentially habitable siblings in multi-planet systems.

Observational efforts are increasing data precision on planetary mass and radius. These efforts must be accompanied by theoretical effort to improve precision in structural and compositional models. Our updated interior models include a more profound understanding of \cdnote[petrology]. Compared with \citet{dorn2017generalized}, the updated model includes (1) liquid rock phases in mantle and core, (2) light core elements, (3) improved equations of state, and most importantly (4) the dissolution of water in the magma ocean depending on solubility equilibria. Future model improvements may include the addition of other volatiles in both interior and atmospheric structure \citep{2021arXiv210812902G} and the dependency of solubility on mantle composition.

For short-period exoplanets inside the runaway greenhouse transition, our results suggest the following. \cdnote[The bulk water budget will be divided between a surface and the magma ocean reservoir. While common interior models include surface reservoirs, the deep magma ocean reservoir is usually neglected with severe consequences for inferred water mass fractions.]
By necessity, any inferred surface water reservoir generally implies an even larger water reservoir in the mantle.
Hence, inferred water mass fractions can vary by up to an order of magnitude depending on whether deep magma ocean water reservoirs are included or not. The differences depend on planet mass, the parameterisation of the thermal structure of a steam atmosphere, as well as the actual water mass fraction.
We have tested three different interior models of various degrees of complexity. For a given interior composition, calculated radii can vary by up to 16\%, \cdnote[which can be well above the precision obtained from transit surveys (e.g., TESS, CHEOPS)].
The potential for a much larger, undetected storage capacity of volatiles inside nominally rocky super-Earths indicates that they may host a larger volatile budget than previously estimated. We will examine this question in more detail in future work.

Exoplanet population studies may \cdnote[be sensitive to the effect of deep water reservoirs. Current data already allow to] infer the fraction of short-period exoplanets in runaway greenhouse states and their total bulk volatile budgets. In summary, our results suggest that volatile inferences from retrievals need to account for the additional mass that can be locked up in the interior in highly molten states. The presence or absence of magma oceans in planetary mantles affects the bulk volatile abundances and climatic conditions of rocky exoplanets to first order. 
Ultimately, deep volatile reservoirs are crucial to consider for a better understanding of exoplanet interiors, with important implications for their formation and evolution.
%
\acknowledgments
We thank an anonymous reviewer for help in improving the manuscript. C.D. was supported by the Swiss National Science Foundation (SNSF) under grant PZ00P2\_174028. T.L. was supported by the Simons Foundation (SCOL Award  No. 611576). This work was in part carried out within the framework of the National Centre for Competence in Research PlanetS supported by the Swiss National Science Foundation and benefited from information exchange within the program ‘Alien Earths’ (NASA grant No. 80NSSC21K0593) for NASA’s Nexus for Exoplanet System Science (NExSS) research coordination network.  

\appendix
\renewcommand\thefigure{\thesection\arabic{figure}}
\setcounter{figure}{0}

\section{Solubility law}

A solubility law relates the abundance of water in the magma to its abundance in the atmosphere, according to the partial pressure of water at the surface of the magma ocean.

\begin{figure}[tbh]
\centering
	\includegraphics[width=.5\linewidth,trim=0cm 0cm 0cm 0cm, clip]{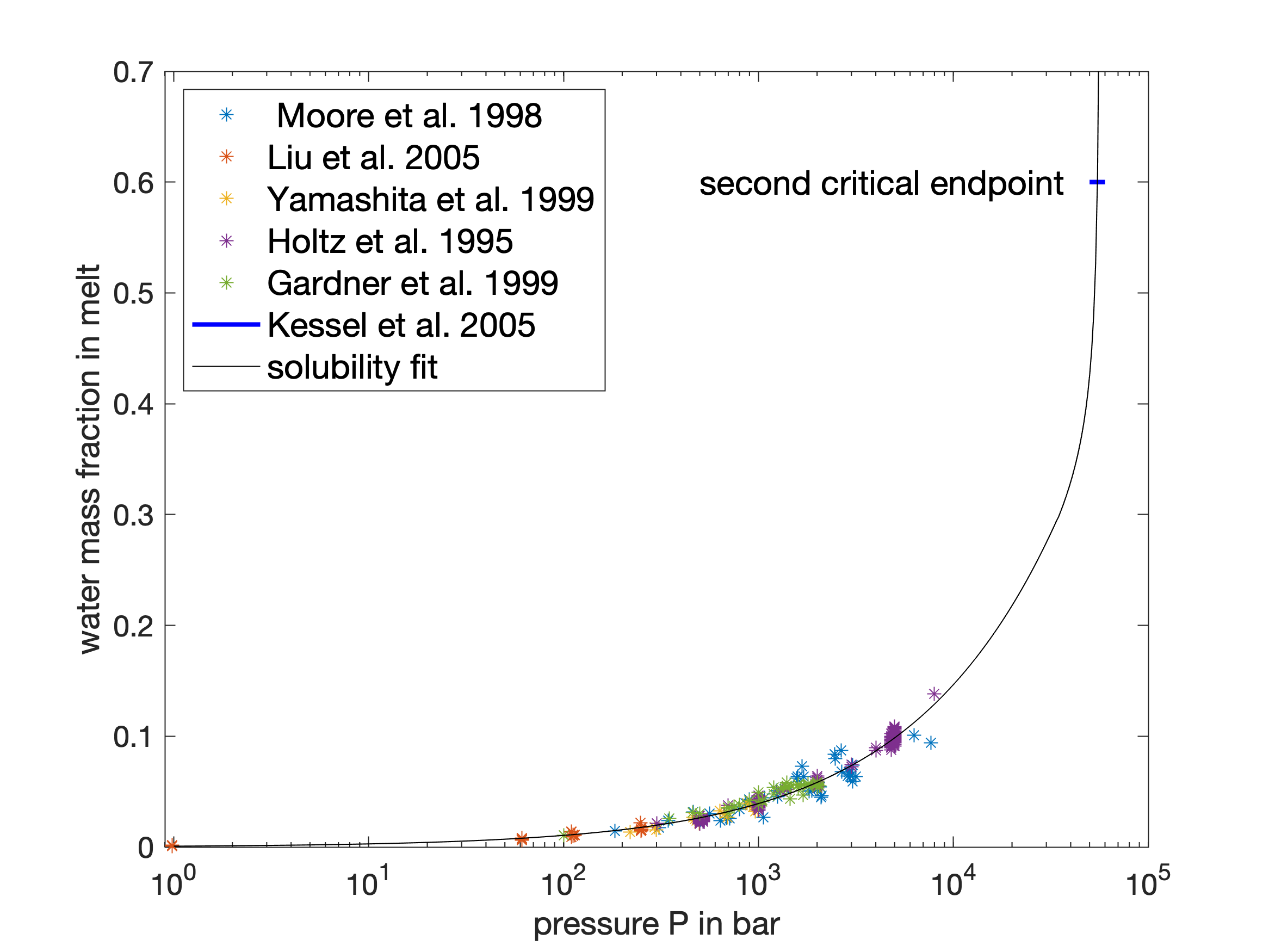}
	\caption{Solubility of water in silicate melts as a function of pressure from different petrological experiments. The fit in solid black is used in our study and connects the region of moderate solubilities ($<1$GPa) to the region above the critical second endpoint where water becomes fully miscible in melt. The solubility curve neglects any effects of melt composition.}
    \label{fig:app}
\end{figure}

 \bibliographystyle{aasjournal}
 \bibliography{library,tl} %

\label{lastpage}

\end{document}